\documentclass[11pt]{article}
\usepackage{latexsym}

\newtheorem{theorem}{Theorem}

\newtheorem{lemma}{Lemma}

\begin{document}

\title{ \bf Boundary Value Problem for $r^2 \,d^2 f/dr^2 + f = f^3$ (II):
 Connection Formula}

\author{Chie  Bing Wang\thanks
  {\small \it Current address:Department of Mathematics, University of California,
  Davis, CA 95616.e-mail:cbwang@math.ucdavis.edu } \\
 \small \it  Department of Mathematics, University of Pittsburgh \\
 \it Pittsburgh, PA 15260.}

\date{}
\maketitle

\begin{abstract}
In this paper, we  study the analytic expansion in a small neighborhood
of $0$ in the complex plane for the solution to
the equation  $p \,dp/dz - p = z(z-1)(z-2)$ satisfying $p(z) = -z + O(z^2)$
as $z \to 0$. We show that the expansion is valid
for $|z| \le s_0$, where $s_0 >1$. Then we get an explicit formula for
$p(1)$ which is used to give the connection formula for
the problem $r^2 f^{''} + f = f^3$, $f(1) = 0, f(\infty) = 1$.
\end{abstract}

%\begin{keywords}
%monopole, shooting method,analytic continuity,
%asymptotics,connection formula
%\end{keywords}

%\begin{AMS}
%34B15, 34B30, 34E05
%\end{AMS}

%\pagestyle{myheadings}
%\thispagestyle{plain}
\markboth{C. B. Wang}{ Monopole in the Pure SU(2) Gauge Theory}

\setcounter{equation}{0}
%%%%%%%%%%%%%%%%%%%%%%%     1     %%%%%%%%%%%%%%%%%%%%%%%%%%%%%%
\section{Introduction}
%\newcounter[subsection]{section}
%\newcounter[equation]{subsection}
%\subsection*{}

  In this paper we continue the discussion in \cite{wang1}
for the boundary value problem
    \begin{eqnarray}
  & &    r^2 f'' + f = f^3, 0<r<\infty,    \label{eq1.1}   \\
 & & f(r)  \to 0,  \,  {\rm as \,\,} r \to 0,  \label{eq1.2}  \\
 & & f(\infty)=1,  \label{eq1.3}
    \end{eqnarray}
where $f=f(r)$, and $'$ means the derivative. By the transformation
$r= e^x,   f(r) = y(x)$, the equation is changed to 
$y'' - y' + y=y^3$. In \cite{wang1}, we proved that 
the following boundary value problem 
   \begin{eqnarray}
   & &  y'' - y' + y=y^3, \,\, 0<x< \infty,  \label{eq3.1}   \\
(P^+)& &y(0) = 0, y(\infty) = 1,  \label{eq3.2}  \\
   & &  y(x) > 0, \,\,\, 0<x< \infty,  \label{eq3.3}
   \end{eqnarray}
has a unique solution $y^*(x)$, and we denote $a^*= y^*{^{'}}(0)$, which is 
a positive number. We will find the value of this number $a^*$, which is
very important for this equation.

  Since $y^*(x)$ is monotonically increasing on
 $x \in [ - \epsilon^-, \infty)$ for some small number $\epsilon^- >0$,
and $y^*(0)=0$, $y^*(\infty)=1$, we have $z(x)=1-y^*(x)$ is ranged in
$[0, 1- y^*(- \epsilon^-) ]$, where $y^*(- \epsilon^-)$ is 
a small negative number. We take $z^{'}(x)$ as a function of
$z$ for $z \in [0, 1- y^*(- \epsilon^-) ]$, say $z^{'} = P(z)$.
Then $P(z)$ satisfies a first order differential equation derived from
the equation for $y^*$, and a corresponding asymptotics as $z \to 0$,
derived from the asymptotics of $y^*$ as $x \to \infty$.
This forms an initial value problem for $P(z)$.
If we can get the analytic formula for $P(z)$, then we 
are at least able to get  $a^*=y^*(0)$  by the property of
$y^*(x)$ at $ x = +\infty$, which is a way to solve the connection problem,
and get the global behaviour of the solution.

 To solve the initial value problem, we change the
the first order equation into an integral equation
combined with the initial condition
   \begin{eqnarray*}
   q(z)^2 &=& -2 \int_0^1 t q(zt) dt+2 -2 z+{z^2 \over 2}, \hspace{1cm}
            0 \le z \le 1+\delta_0,     \\
   q(z) &=& 1+ O(z), z \to 0,  
    \end{eqnarray*}
where $\delta_0 = -y^*(- \epsilon^-)>0 $ is a small number. 
We first show that this problem 
has a unique solution. We construct the solution by an iteration
sequence $\{ q_n(z) \}$, which gives 
$P(z) = -z \lim_{n \to \infty} q_n(z)$ for $z \in [0, \epsilon]$
for a small $\epsilon >0$.

   We then want to analytically extend the solution to a neighborhood
of $0$ in the complex plane $(|z|\le \epsilon)$. 
We first extend $\{ q_n(z) \}$ into $|z| \le\epsilon$, which is seen
by the construction of $\{ q_n(z) \}$.  
The $\{ q_n(z) \}$ form a uniformly bounded sequence on 
$|z| \le \epsilon$. By the normal family theory \cite{ahlfors},
 $\{ q_n(z) \}$ has 
a convergent subsequence in $|z| \le \epsilon$, which
converges to an analytic function in $|z| \le \epsilon$.
Specially this analytic function is equal to $P(z)$
for $z \in [0, \epsilon]$. So we have extend $P(z)$ from
the interval $ [0, \epsilon]$ to $|z| \le \epsilon$. 
Therefore we get the analytic expansion 
$P(z)=\sum_{n=1}^{\infty} b_n z^n$ for $|z| \le \epsilon$,
 where the coefficients 
$b_n$ can be obtained from the equation of $P(z)$.
We then show this series is convergent to $P(z)$
wherever $P(z)$ is bounded by the fact that 
this series is convergent to $P(z)$ on $[0, \epsilon]$, and
$P(z)$ is bounded on $[0, 1+\delta_0]$. 
We then obtain a formula for $a^*=y^*(0)$.
The analytic continuity to a small neighborhood of $0$
in complex plane will be done in Sect.~2, 
and the convergent radius will be discussed in Sect.~3.
This method also works for other similar equations to solve
the connection problem.

\setcounter{equation}{0}
%%%%%%%%%%%%%%%%%%%%%%%%     4      %%%%%%%%%%%%%%%%%%%%%%%%%%%%%%%
\section{Analytic Continuity}

  It  has been proved that the problem $(P^+ )$ has a unique solution
     \[ y^*(x) =y(x,a^*). \] 
We then want to extend the solution to the negative
axis. The method we will use to extend the solution is tightly
related to the value of $a^*$. So we need more properties of the 
solution. The idea here is to represent ${y^*}'$ as a function
of $y^*$, and then to reduce (\ref{eq3.1}) to a first-order
equation. Then the first-order equation might help us to get
the value of $a^*$.

   Since ${y^*}'(0)=a^* >0$, we can choose a small $\epsilon^- > 0$,
such that ${y^*}'(x) >0$,
for $x \in [-\epsilon^-, \infty)$ (Lemma 4 \cite{wang1}).  Let
   $ \delta_0 = - y^* (-\epsilon^-) >0$, and
\begin{equation}
z(x) = 1 - y^*(x),  \label{addeq4.1}
\end{equation}
for $-\epsilon^- <x < \infty$.
Then  equation (\ref{eq3.1}) becomes
   \begin{equation}
   z'' - z' = z(z-1)(z-2),
   \end{equation}
where $0 < z \le 1 + \delta_0$, and $z(x) \to 0$, as
$x \to \infty$.
Because $z'$ is always negative(Lemma 4 \cite{wang1}),i.e., $z(x)$ is
strictly monotone, $z'(x)$ can be thought as a
function of $z$. Let
    \begin{equation}
    z'(x) =  P(z(x)).    \label{eq4.3}
    \end{equation}
We then obtain the following lemma by Lemma 4 (iii) \cite{wang1}.
\begin{lemma}   \label{Lemma4.1}
 $p=P(z)$ solves the following problem
   \begin{eqnarray}
  & &  p\, p' - p = z(z-1)(z-2), 0<z<1+\delta_0,  \label{eq4.5}   \\
  & &    p(z) = -z + O(z^2), z \to 0,  \label{eq4.6}
   \end{eqnarray}
and
   \begin{eqnarray}
      a^* &=& - P(1),  \label{addeq4.7} \\
      {a^*}^2 &=& {1 \over 2} + 2 \int_0^1 P(z) \, dz. \label{addeq4.8}
   \end{eqnarray}
\end{lemma}
\noindent {\it Proof.}
In Lemma 4 (iii) \cite{wang1}, we obtained that
$ { y'(x) \over y(x) - 1 } = -1 + O(e^{-x})$, $y(x) = 1 - c e^{-x} + O(e^{-2x})$,
    as $x \to \infty$, which implies (\ref{eq4.6}).
The formula (\ref{addeq4.7}) is obtained from
$a^* = y^{*^{'} }(0)$, by using (\ref{addeq4.1}), (\ref{eq4.3})
and the fact $y^*(0)=0$.
And (\ref{addeq4.8}) follows from taking integral from $0$ to $1$ 
on both sides of 
the equation (\ref{eq4.5}) and using (\ref{addeq4.7}).
\,\,\,\, $\Box$

     We next want to find analytic expression of $P(z)$, so
that $a^*=- P(1)$ can be explicitly defined. The plan is as follows.
First we use the integral equation obtained from 
(\ref{eq4.5}) to express $P(x)$ as a limit
of a function sequence in a neighborhood of $0$. Each function
of the sequence can be analytically extended to a neighborhood
of $0$ in the complex plane. Then by the normal family theory of
analytic functions, this sequence has a convergent subsequence,
which converges to $P(z)$ on the real axis part, and converges
to an analytic function in the neighborhood of $0$ in the
complex plane. So $P(z)$ is analytically continued to a
neighborhood of $0$ in the complex plane.

    Then we have the Taylor's series expansion of $P(z)$
in this neighborhood. We then prove the convergent
radius of this series is greater than $1$, so that the
expression of $a^*$ is found, which will be discussed in
the next section.

    Equation (\ref{eq4.5}) with the condition (\ref{eq4.6}) is
equivalent to the following integral equation
   \begin{equation}
    {1 \over 2} p^2 = \int_0^z p(s) \, ds
       + z^2 -z^3+{z^4 \over 4}, \hspace{1cm} 0 \le z \le 1+\delta_0.
                      \label{eq4.10}   
   \end{equation}
Let
      \[ p(z) = - z q(z).  \]
Then  (\ref{eq4.10}) becomes 
    \begin{equation}
  q(z)^2 = -2 \int_0^1 t\, q(zt)\, dt+2 -2 z+{z^2 \over 2},\hspace{1cm}
            0 \le z \le 1+\delta_0. \label{eq4.15}  
    \end{equation}

\begin{lemma}   \label{Lemma4.2}
The following problem
    \begin{eqnarray}
   q(z)^2 &=& -2 \int_0^1 t \,q(zt) \,dt+2 -2 z+{z^2 \over 2},\hspace{1cm}
            0 \le z \le 1+\delta_0,   \label{eq4.16}  \\
   q(z) &=& 1+ O(z), z \to 0,  \label{eq4.17} 
    \end{eqnarray}
has a unique solution.
And then we obtain that equation (\ref{eq4.5}) with (\ref{eq4.6}) has
unique solution $P(z)$.
\end{lemma}

\noindent {\it Proof.}
We have seen that this problem has a solution by Lemma 1.
Now suppose there are two solutions $q(z)$ and $\bar{q}(z)$
to the problem. Then by (\ref{eq4.17}) there exist 
$M_1 >0, \, \epsilon_1 >0$, satisfying $M_1 \epsilon_1 <{2 \over3}$,
so that if $0 \le z \le \epsilon_1$,
   \begin{eqnarray}
   |q(z)-1| &\le& M_1 z \le M_1 \epsilon_1 <{2 \over3},
              \label{eq4.18}  \\
   |\bar{q}(z)-1| &\le& M_1 z \le M_1 \epsilon_1 <{2 \over3},
              \label{eq4.19}
   \end{eqnarray}
which imply that
     \[ q(z), \bar{q(z)} > 0.  \]
Since $q, \bar{q}$ satisfy equation  (\ref{eq4.16}), there is
   \[
    (q(z) +\bar{q}(z)) |q(z)-\bar{q}(z)|
        \le 2 \int_0^1 t \left( |q(zt)-1|+|\bar{q}(zt)-1|
            \right) \, dt   
        \le {4\over3} M_1 |z|.  \]
By (\ref{eq4.18}) and (\ref{eq4.19}), we have
   \[  q(z)+ \bar{q}(z) \ge 2 ( 1 - M_1 \epsilon_1).  \]
Therefore we arrive at
    \[  |q(z) - \bar{q}(z)| \le 2 M_1 \gamma_1 |z|,  \]
for $0 \le z \le \epsilon_1$, where
$\gamma_1={1 \over 3(1-M_1 \epsilon_1) }$. By induction
method(see the proof of Lemma 4   ), we can prove that
     \[ |q(z) - \bar{q}(z)| \le 2 M_1 \gamma_1^n |z|,  \]
for any integer $n>0$, and $0 \le z \le \epsilon_1$. Letting
$n \to \infty$, we see that $q(z) = \bar{q}(z)$
for $0 \le z \le \epsilon_1$.
    
    When $z > \epsilon_1$, consider equation  (\ref{eq4.5}).
Since $P(z) < 0$, for $z \in [\epsilon_1,1+\delta_0]$,
the Lipschitz condition is satisfied. Then the general uniqueness
theorem \cite{coddington} of solution implies $q=\bar{q}$.  \,\,\,\, $\Box$

   Now let us make an iteration sequence
    \begin{eqnarray}
    q_0(z) &=& 1,       \label{eq4.20}   \\
    q_{n+1}^2 (z) &=& -2\int_0^1 t\, q_n(z t) \,dt+2-2 z+{z^2 \over 2},
                \hspace{1cm}     n \ge 0,   \label{eq4.21}
    \end{eqnarray}
in order to give a limit representation for the solution $q(z)$.
%%%%%%%%%%%%%%%  define M, and epsilon
Choose $\epsilon_0, M >0$ satisfying
   \begin{eqnarray*}
   & & M \, \epsilon_0 < {2 \over 3},  \\
   & & (2 + {\epsilon_0 \over 2}) ( 1 + M \epsilon_0) \le M,    \\
   & & (1+M \epsilon_0) ({2\over 3} M +2+{\epsilon_0 \over 2} ) \le M.
   \end{eqnarray*}
For example we can choose $M=10$. Then a sufficiently small
$\epsilon_0$ would satisfies all of the three conditions. 
We have the following results for the iteration sequence.

\begin{lemma}    \label{Lemma4.3}
   \begin{eqnarray}
   & & q_n(z) >0, \,\,\,\, n \ge 0,  \\
   & & | q_n(z) - 1 | \le M |z| < 1, \,\,\, n \ge 1,  \label{eq4.23}
   \end{eqnarray}
for $ 0 \le z \le \epsilon$, where
    \[ \epsilon = min \left( {1 \over 4}, \epsilon_0 \right). \]
\end{lemma}

\noindent {\it Proof.}
By (\ref{eq4.20}) and (\ref{eq4.21}), there is
    \begin{equation}
    q_1^2 = 1 - 2 z + {z^2 \over 2}.  \label{eq4.25}
    \end{equation}
Let
  \[ f_1(x) = -2 z + {z^2 \over 2},  \]
which satisfies
   \[ |f_1| \le 2 \epsilon + {\epsilon^2 \over 2} < 1,  \]
for $0 \le z \le \epsilon$, since $\epsilon < 1/4$.
  We then have
     \begin{eqnarray*}
     |q_1 - 1| &=& | (1+f_1)^{1 \over 2} -1 |  \\
       &=& \arrowvert \sum_{m=0}^{\infty}
         {{1\over2}({1\over2}-1)\dots({1\over2}-m+1) \over m! }
          f_1^m - 1 \arrowvert   \\
       &\le& \sum_{m=1}^{\infty} |f_1|^m
          ={ |f_1| \over 1 - |f_1| },
     \end{eqnarray*}
where we have used $|{1\over 2} -j |= j-{1\over 2} < j \,\, (j>0)$,
and $|({1\over2}-1)\dots({1\over2}-m+1)| < (m-1)!\,\, (m>1)$.
Since $0< \epsilon < \epsilon_0 $, by the definition of 
$M$ and $\epsilon_0$ above, we have
   \[ (2 + {\epsilon \over 2}) ( 1 + M \epsilon) \le M,  \]
which implies  
$ | f_1 (z) | ( 1 + M |z| ) \le ( 2 |z| + {|z|^2 \over 2})(1+ M |z|)
\le |z| (2 + {\epsilon \over 2}) ( 1 + M \epsilon) \le M |z| $,
or $| f_1 (z) | \le ( 1 - | f_1 (z) | ) M |z|$,
and then
   \[ { |f_1| \over 1 - |f_1| } \le M |z|.  \]
Hence $ | q_1 - 1| \le M |z|  $.

    Now suppose  $ | q_n -1 | \le M |z| $.   
Let us show  $ | q_{n+1} - 1 | \le M |z| $.

   Let
     \[ f_{n+1}=-2 \int_0^1 t ( q_n(z t) -1 ) dt
            - 2 z + {z^2 \over 2}.  \]
Then we have
   \[ q_{n+1}^2 = 1 + f_{n+1}.   \]
We want to show
    \begin{equation}
     |f_{n+1} | \le { M |z| \over 1+ M |z| }.  \label{eq4.30}
    \end{equation}
In fact,  there is
    \begin{eqnarray*}
    |f_{n+1}| &\le& 2 \int_0^1 t M |z| t dt
           + 2 |z| + {|z|^2 \over 2}   \\
       &=& {2\over3} M |z| + 2 |z| + {|z|^2 \over 2}   \\
       &\le& ({2\over 3} M +2+{\epsilon \over 2} ) |z|.
    \end{eqnarray*}
Now because $0<\epsilon \le \epsilon_0$, by the definition of
$\epsilon_0$, we deduce
   \begin{eqnarray*}
  (1+ M |z| ) |f_{n+1}| &\le& ( 1+ M \epsilon) |f_{n+1} | \\
   &\le& (1+M \epsilon)
          ({2\over 3} M +2+{\epsilon \over 2} ) |z|    \\
   &\le& (1+M \epsilon_0)
          ({2\over 3} M +2+{\epsilon_0 \over 2} ) |z|    \\
   &\le& M |z|,
   \end{eqnarray*}
which implies  (\ref{eq4.30}), and then
    \begin{equation}
    { |f_{n+1}| \over 1 - |f_{n+1}| } \le M |z|.
    \end{equation}
So we get
    \begin{eqnarray*}
    |q_{n+1} - 1|
    &=& \arrowvert (1+f_{n+1})^{1 \over 2} -1 \arrowvert \\
    &=& \arrowvert \sum_{m=0}^{\infty}
    {{1\over2}({1\over2}-1)\dots({1\over2}-m+1) \over m! }
    f_{n+1}^m - 1 \arrowvert   \\
    &\le& { |f_{n+1}| \over 1 - |f_{n+1}| } \le M |z|.
    \end{eqnarray*}
The lemma is proved.   \,\,\,\, $\Box$

\begin{lemma}   \label{Lemma4.4}
If $0 \le z \le \epsilon$, where $\epsilon$ is defined in Lemma 3, then  there is
   \begin{equation}
   | q_{n+1} - q_n | \le M \gamma^n |z|,  \label{eq4.40}
   \end{equation}
for $n \ge 0$, where
  \[ \gamma = { 1 \over 3 (1 - M \epsilon) } < 1.
  \]
\end{lemma}

\noindent {\it Proof.}
Since $q_0=1$, by Lemma 3    , we see that
   \[ | q_1 - q_0 | \le M |z|.   \]
Now suppose
   \[   | q_{n} - q_{n-1} | \le M \gamma^{n-1} |z|,  \]
let us show equation  (\ref{eq4.40}) is true.

   In fact, when $n \ge 1$,  there is
   \begin{eqnarray*}
  (q_{n+1} + q_n) \left| (q_{n+1} - q_n) \right|
 &=&  \left| (q_{n+1} + q_n) (q_{n+1} - q_n) \right|   \\
 &=& \left| -2 \int_0^1 t (q_n(z t) - q_{n-1}(z t) ) dt \right| \\
 &\le& 2 \int_0^1 t \left| q_n(z t) - q_{n-1}(z t) \right| dt \\
 &\le& {2 \over 3} M \gamma^{n-1} |z|.
   \end{eqnarray*}
Still by Lemma 3    , we have
    \[ q_{n+1} + q_n \ge 2 - 2 M |z| \ge 2 - 2 M \epsilon.  \]
Thus
   \begin{eqnarray*}
    \left| q_{n+1} - q_n \right|
   &\le& { 2 \over 6(1 - M \epsilon) } M \gamma^{n-1} |z|  \\
   &=& M \gamma^n |z|.
   \end{eqnarray*}
So we have proved the lemma. \,\,\,\, $\Box$

\begin{lemma}   \label{Lemma4.5}
For the $P(z)$ defined by (\ref{eq4.3}), we have
  \begin{equation}
    P(z) = -z \lim_{n \to \infty} q_n(z),  \label{eq4.45} \\
  \end{equation}
for $0 \le z \le \epsilon$.
\end{lemma}

\noindent {\it Proof.}
Since
    \[ q_{n+1}(z) = q_0 (z)+ \sum_{k=0}^{n}
          ( q_{k+1}(z) - q_n (z) ),  \]
by Lemma 4    , we see that the sequence ${q_n(z)}$ is
uniformly convergent to a function, say $q(z)$ on
$[0, \epsilon]$. $q(z)$ satisfies  (\ref{eq4.16})  and
(\ref{eq4.17} ) for $0 \le z \le \epsilon$. 
By Lemma 2, $q(z)$ is the unique solution.
Thus $P(z) = - z q(z)= -z \lim_{n \to \infty} q_n(z)$. \,\,\,\, $\Box$

   Define
     \[ B_{\epsilon} = \{ z \in {\bf C} \mid \, |z| \le \epsilon \},
     \]
where ${\bf C}$ is the complex plane.
\begin{theorem}    \label{Theorem4.6}
There is unique analytic continuation of $P(z)$ 
in $B_{\epsilon}$, and then $P(z)$ has the series expansion
  \begin{equation}
    P(z) = \sum_{n=1}^{\infty} b_n z^n,  \label{eq4.50}
  \end{equation}
for $z \in  B_{\epsilon}$.
\end{theorem}

\noindent {\it Proof.}
Let us go back to the sequence $\{ q_n(z) \}$ defined by
(\ref{eq4.20}) and (\ref{eq4.21}) for $0 \le z \le \epsilon$.
We claim that the sequence can be analytically continued to 
$ z \in B_\epsilon $. In fact, if we review the proof of
Lemma 3    , it not hard to find that (\ref{eq4.23}) is true
for $z \in B_\epsilon$. So each $q_n(z)$ is well defined and
analytic in $B_\epsilon$.
And the sequence  $\{ q_n(z) \}$ is uniformly bounded
on $B_\epsilon$. Thus $\{ q_n(z) \}$ is a normal sequence
(or normal family)\cite{ahlfors}. Therefore $\{ q_n(z) \}$
contains an uniformly
convergent subsequence, say $\{ q_{n_k}(z) \}_{k=1}^{\infty}$. 
Let  
  \[ \hat{q}(z) = \lim_{k \to \infty} q_{n_k}(z), \,\, z \in B_\epsilon. \]
Specifically by Lemma 5    
  \[ \hat{q}(z) = \lim_{k \to \infty} q_{n_k}(z)
     = \lim_{n \to \infty} q_{n}(z) = q(z), \,\, 0 \le z \le \epsilon.  \]
Hence  $P(z)$ is analytically extended into
$B_\epsilon$, and $P(z)$ has the Taylor expansion (\ref{eq4.50}),
where the coefficients $b_n$ are uniquely defined by the equation
(\ref{eq4.5}) and (\ref{eq4.6}). If $P(z)$ has another extension,
then the corresponding  series expansion restricted on $[0, \epsilon]$
is the series (\ref{eq4.50}) since $P(z)$ must satisfy the equation
 (\ref{eq4.5}) and (\ref{eq4.6}) on $[0, \epsilon]$.
That means it is still the same series.
     \,\,\,\, $\Box$

    By Theorem 1  and (\ref{eq4.3}) we see that $y^*(x)$ satisfies
the following first-order equation
   \[ 
     y'+ b_1 (1-y) + b_2 (1-y)^2 + \cdots + b_n (1-y)^n + \cdots =0, 
   \]
at least for $x > 0$.

\setcounter{equation}{0}
%%%%%%%%%%%%%%%%%%%%%%%%     5      %%%%%%%%%%%%%%%%%%%%%%%%%%%%%%%
\section{The Value of $a^*$}
   In the last section we have seen that $P(z)$, as an analytic function 
in the complex domain, has the series expansion. In this section, we
investigate this series, and further get the formula for $a^*$.

\begin{lemma}     \label{Lemma5.1}
There is the following recursion relation for the
coefficients $b_n$ in (\ref{eq4.50})
   \begin{eqnarray}
   b_1 &=& -1,   \\
   b_2 &=& {3 \over4},  \\
   b_3 &=& {1 \over 40}  ,  \\
   b_n &=&  {1 \over n+2} \sum_{k=2}^{n-1} k \, b_k \, b_{n-k+1},
   \end{eqnarray}
where $n \ge 4$. And
   \[ b_n > 0 \]
for $n \ge 2$.
\end{lemma}

\noindent {\it Proof.}
Because (\ref{eq4.50}) is the solution to  (\ref{eq4.5}), 
substitution of the series in (\ref{eq4.50}) into the 
equation (\ref{eq4.5}) we then get the recursion
relation by comparing the coefficients of $z^n$ on
both sides of the equation. And it's easy to see that
$b_n > 0$ for $n \ge 2$.  \,\,\,\, $\Box$

\begin{lemma}      \label{Lemma5.2}
Let $r_0$ be the convergent radius of series (\ref{eq4.50}). We have
    \begin{equation}
         r_0 \ge 1 + \delta_0.  \label{eq5.5}  \\
    \end{equation}
\end{lemma}

\noindent {\it Proof.}
By Theorem 1 , we see that
    \[ r_0 > 0.  \]
If $r_0 < 1 + \delta_0$, then there are two possibilities.

   (i) $\sum_{n=1}^{\infty} b_n r_0^n $ is not convergent.
Because $b_n > 0(n \ge 2)$, we then have
    \begin{equation}
      \sum_{n=1}^{\infty} b_n r_0^n = + \infty.  \label{eq5.20}
    \end{equation}
Let
   \[ R(z) = z + P(z).   \]
Since $P(z)$ is bounded, $R(z)$ is also bounded. Then
$b_n > 0(n \ge 2)$ implies
    \[ 0 < \sup_{[0, r_0]} R(z) < \infty.  \]
Now choose   a positive number
     \[  R_0 > \sup_{[0, r_0]} R(z).  \]
Set
    \[  s_N (z) = \sum_{n=2}^{N} b_n z^n.   \]
By (\ref{eq5.20}), there is a $N>0$, such that
   \[  s_N (r_0) > R_0.   \]
Since $s_N (z)$ is a polynomial, which is continuous at
$z = r_0$, there is $\delta_1 \in (0,r_0)$, such that
if $\delta \in [0, \delta_1]$, there is
  \[ s_N (r_0 - \delta) > R_0,    \]
which implies
  \[ R(r_0-\delta) > s_N (r_0 - \delta) > R_0
         > \sup_{[0, r_0]} R(z).    \]
This is a contradiction.

    (ii) $\sum_{n=1}^{\infty} b_n r_0^n $  is convergent,
but $\sum_{n=1}^{\infty} b_n (r_0+\delta)^n $ is not
convergent for $\delta > 0$, i.e.
  \[ \sum_{n=1}^{\infty} b_n (r_0+\delta)^n = \infty.  \]
Then there is $N>0$, and a positive sequence $\{\delta_k\}$,
with $\delta_k \to 0$, as $k \to \infty$, such that
    \[ s_N (r_0 + \delta_k) > R_1  \]
for some positive number
    \[  R_1 > \sup_{[0, r_0]} R(z).   \]
Because $s_N (z) $ is continuous, letting $k \to \infty$,
we get
   \[ R(r_0) > s_N (r_0) > \sup_{[0, r_0]} R(z),   \]
which is a contradiction. So we must have
$ r_0 \ge 1 + \delta_0$.   \,\,\,\, $\Box$

\begin{theorem}   \label{Theorem5.3}
We have the following formulas for $a^*$,
  \begin{eqnarray}
  (i) & & a^* = - \sum_{n=1}^{\infty} b_n,  \label{eq5.30} \\
  (ii) & & a^* = \left( {1\over 2} +
    2 \sum_{n=1}^{\infty} {b_n \over n+1} \right)^{1\over2},
                   \label{eq5.31}  \\
  (iii) & & \left( {1\over 2} +
    2 \sum_{n=1}^{N_1} {b_n \over n+1} \right)^{1\over2}
    < a^* <  - \sum_{n=1}^{N_2} b_n,
                   \label{eq5.32}  
  \end{eqnarray}
for any $N_1 \ge 2$, and $N_2 \ge 1$. And specially
    \begin{equation}
     a^* < {1 \over 4}.  \label{eq5.33}
    \end{equation}
\end{theorem}

\noindent {\it Proof.}
  By Lemma 1 and Lemma 7    ,  we obtain (\ref{eq5.30})
and (\ref{eq5.31}). Since $b_n > 0$, for $n \ge 2$,
(\ref{eq5.32}) is true. Choose $N_2 =2$,
(\ref{eq5.33}) is derived.  \,\,\,\, $\Box$

%%%%%%%%%%%%%%%%%%%%%%%%%%%%%%%%%%%%%%%%%%%%%%%%%%%%%%%%%%%%%%%%%%%


\begin{thebibliography}{99} 
\bibitem{ahlfors}
	Ahlfors, L. V.: {\em Complex Analysis},
	 McGraw-Hill, New York, 1979.



\bibitem{coddington}
  Coddington, E.A., N. Levinson, N.:
   {\em  Theory of Ordinary Differential Equations},
     McGraw-Hill,New York, 1955.

%\bibitem{perko}
%  Perko, L.:
%  {\em Differential Equations and Dynamical Systems},
%     2nd ed. Springer-Verlag, New York, 1996.



\bibitem{wang1}
  Wang, C. B.: {\em Boundary value problem for $r^2 \,{d^2 f/dr^2}  + f = f^3$ (I):
  existence and uniqueness}, preprint.

\bibitem{wang3}
  Wang, C. B.: {\em Boundary Value Problem for $r^2 \,{d^2 f/dr^2}  + f = f^3$(III):
   global solution and asymptotics}, preprint.



\end{thebibliography}
\end{document}